\PassOptionsToPackage{unicode}{hyperref}
\PassOptionsToPackage{hyphens}{url}
\PassOptionsToPackage{dvipsnames,svgnames,x11names}{xcolor}
\documentclass[
]{article}
\usepackage{amsmath,amssymb}
\usepackage{lmodern}
\usepackage{iftex}
\ifPDFTeX
  \usepackage[T1]{fontenc}
  \usepackage[utf8]{inputenc}
  \usepackage{textcomp} 
\else 
  \usepackage{unicode-math}
  \defaultfontfeatures{Scale=MatchLowercase}
  \defaultfontfeatures[\rmfamily]{Ligatures=TeX,Scale=1}
\fi
\IfFileExists{upquote.sty}{\usepackage{upquote}}{}
\IfFileExists{microtype.sty}{
  \usepackage[]{microtype}
  \UseMicrotypeSet[protrusion]{basicmath} 
}{}
\makeatletter
\@ifundefined{KOMAClassName}{
  \IfFileExists{parskip.sty}{%
    \usepackage{parskip}
  }{
    \setlength{\parindent}{0pt}
    \setlength{\parskip}{6pt plus 2pt minus 1pt}}
}{
  \KOMAoptions{parskip=half}}
\makeatother
\usepackage{xcolor}
\usepackage[margin=1in]{geometry}
\usepackage{longtable,booktabs,array}
\usepackage{calc} 
\usepackage{etoolbox}
\makeatletter
\patchcmd\longtable{\par}{\if@noskipsec\mbox{}\fi\par}{}{}
\makeatother
\IfFileExists{footnotehyper.sty}{\usepackage{footnotehyper}}{\usepackage{footnote}}
\makesavenoteenv{longtable}
\usepackage{graphicx}
\makeatletter
\def\maxwidth{\ifdim\Gin@nat@width>\linewidth\linewidth\else\Gin@nat@width\fi}
\def\maxheight{\ifdim\Gin@nat@height>\textheight\textheight\else\Gin@nat@height\fi}
\makeatother
\setkeys{Gin}{width=\maxwidth,height=\maxheight,keepaspectratio}
\makeatletter
\def\fps@figure{htbp}
\makeatother
\newlength{\cslhangindent}
\newlength{\csllabelwidth}
\newlength{\cslentryspacingunit} 
\newenvironment{CSLReferences}[2] 
 {
  \setlength{\parindent}{0pt}
  \ifodd #1
  \let\oldpar\par
  \def\par{\hangindent=\cslhangindent\oldpar}
  \fi
  \setlength{\parskip}{#2\cslentryspacingunit}
 }%
 {}
\usepackage{calc}

\usepackage[utf8]{inputenc}
\usepackage[margin = 1in]{geometry}
\usepackage{tabularx}
\usepackage{booktabs}
\usepackage{graphicx}
\usepackage{apacite}
\usepackage{url}
\usepackage{amstext}
\usepackage[justification=raggedright]{caption}
\usepackage[flushleft]{threeparttable}
\usepackage{longtable}
\usepackage{rotating}
\usepackage{setspace}
\usepackage{float}
\ifLuaTeX
  \usepackage{selnolig}  
\fi
\IfFileExists{bookmark.sty}{\usepackage{bookmark}}{\usepackage{hyperref}}
\IfFileExists{xurl.sty}{\usepackage{xurl}}{} 
\hypersetup{
  pdftitle={A Comparison of Reproducibility Guidelines and Its Implications on Undergraduate Statistical Education},
  pdfauthor={Siqi Zheng},
  colorlinks=true,
  linkcolor={blue},
  filecolor={Maroon},
  citecolor={Blue},
  urlcolor={Blue},
  pdfcreator={LaTeX via pandoc}}

\title{A Comparison of Reproducibility Guidelines and Its Implications on Undergraduate Statistical Education}
\usepackage{etoolbox}
\makeatletter
\providecommand{\subtitle}[1]{
  \apptocmd{\@title}{\par {\large #1 \par}}{}{}
}
\makeatother
\subtitle{A Case Study in Bayesian Educational Research}
\author{Siqi Zheng}
\date{28/10/2022}

\begin{document}
\maketitle

{
\hypersetup{linkcolor=}
\setcounter{tocdepth}{2}
\tableofcontents
}
\hypertarget{rethinking-reproducible-research-guidelines}{%
\section{Rethinking Reproducible Research Guidelines}\label{rethinking-reproducible-research-guidelines}}

Reproducibility of scientific findings is the core of scientific research (\protect\hyperlink{ref-tetens-2016}{Tetens, 2016}). Scientists have proposed recommendations and suggested criterion (\protect\hyperlink{ref-chang-2015}{Chang \& Li, 2015}; \protect\hyperlink{ref-yale-2010}{Yale, 2010}) for better scientific reproducibility, but the reproduction of actual scientific research may still be challenging even if the research article meets the expectations. In this project, we replicated a research article by professor Skinner (\protect\hyperlink{ref-skinner-2019A}{2019a}), which explores the association between broadband access and online course enrollment in the US. We summarized the model results and key findings from our replication and compared our results with the original project. Based on the replication experience, we aim to demonstrate the unexpected challenges of statistical research reproduction, even when codes and data are shared openly and the quality of the materials for reproducibility on GitHub (\protect\hyperlink{ref-skinner-2019B}{Skinner, 2019b}) are relatively high. Moreover, we investigate the implicit presumptions of the researchers' level of knowledge and discuss how such presumptions may add difficulty to the reproduction of scientific research. Finally, we hope this article sheds light on the design of reproducibility criterion and opens up a space to explore what should be taught in undergraduate statistics education.

\hypertarget{background-of-the-case-study}{%
\subsection{Background of the Case Study}\label{background-of-the-case-study}}

Online courses can be a solution to accessible education during the COVID-19 pandemic. Given the benefits of remote studying, it is important for the universities to understand how internet access affects online course enrollment. Professor Benjamin T. Skinner utilized data from the National Broadband Map and the Integrated Postsecondary Education Data System (IPEDS) to investigate the relationship between various measures of broadband access (download speed, upload speed, and the number of providers) and the number of students who take undergraduate online courses at public colleges and universities with open admissions policies using Bayesian regression models (\protect\hyperlink{ref-skinner-2019A}{Skinner, 2019a}). The original paper shows that increases in broadband speed are positively correlated with the number of students who take online courses given a relatively lower internet speed, but the rate of increase in number of students decline as internet speeds increase. These findings may suggest a possible threshold for minimal internet speed and are particularly helpful for students in communities with low average broadband speeds. During the COVID-19 pandemic, many students, including those in the communities with less access to the internet, are forced to attend online courses. If re-examining this paper produces the same results, we may want to rethink how the government can better allocate internet resources for students and how schools nowadays can support these students in an online class. In our paper, we would replicate the models on our local computer with online data according to the instructions on GitHub (\protect\hyperlink{ref-skinner-2019B}{Skinner, 2019b}).

\hypertarget{a-literature-review-of-reproducible-recommendationsguidelines}{%
\subsection{A Literature Review of Reproducible Recommendations/Guidelines}\label{a-literature-review-of-reproducible-recommendationsguidelines}}

A review of past guidelines of reproducible research may reveal the essence of reproducibility for us before diving into the actual replication. These guidelines may help us assess and define the strength of the original academic paper as well. Thus, we first compare the original project against some general guidelines, particularly the comprehensive ones by Yale Law School Roundtable on Data and Code Sharing (\protect\hyperlink{ref-yale-2010}{Yale, 2010}) and a group of researchers who reviewed 60 economic papers (\protect\hyperlink{ref-chang-2015}{Chang \& Li, 2015}). Then we determine its reproducibility according to a discipline-specific guideline by the National Science Foundation (NSF) and the Institute of Education Sciences, U.S. Department of Education (IES) (\protect\hyperlink{ref-national-science-foundation-2018}{National Science Foundation \& Institute of Education Sciences, U.S. Department of Education, 2018}). In 2006, Wicherts et al. (\protect\hyperlink{ref-wicherts-2006}{2006}) provided a simple solution to scientists for better scientific reproducibility: report data and standardized codes publicly in appendix. This solution pinpoints the core idea of reproducible research: the technical tools and raw materials for a research project.

4 years later, researchers from Yale Law School Roundtable suggested a more complete list of 6 general recommendations for scientists to produce reproducible results (\protect\hyperlink{ref-yale-2010}{Yale, 2010}). The first step is to provide scripts and the data as much as possible. This is a common basic requirement for reproducible scientific research. In our case, the author provides full codes with limited data due to legal reasons. For the unattached datasets, the author includes codes to download the data and provides links to all data in the Readme file. In addition, the author even offers a GNU makefile for one-step replication (\protect\hyperlink{ref-skinner-2019B}{Skinner, 2019b}). The second recommendation is to create a unique identifier to each version of the released code. Since there is only one version for my case, no identifier is used. Nonetheless, the author recorded the date of his last successful attempt with the makefile. The third suggestion is about the ``computing environment and software version used in the publication, with stable links to the accompanying code and data.'' A virtual machine would be even better. In our case, the professor includes the computing environment (MacOS) and software applications (including their official websites) in the Readme file. The author does not, however, specify the versions of the software applications and R packages. The fourth, fifth and sixth recommendations are about using open licensing, open access to the journal article and the preprints, and nonproprietary formats for legal reproduction. Our case satisfies all conditions. Based on these recommendations, the reproducibility of this research project should be very high.

5 years later, Chang \& Li (\protect\hyperlink{ref-chang-2015}{2015}) provided a more comprehensive list of general recommendations based on their review of 60 economic research papers. A close comparison with the previous papers shows that 5 extra recommendations to researchers are added in this recent paper. First, they believe authors should report the expected model estimation time (especially for Bayesian models). Based on my past experience with stan, I believe it is hard to define the expected time requirement given that the running time of stan (a C++ library used for Bayesian computation) depends on the computing environment and the types of computing architecture. Our case, which utilizes a Bayesian hierarchy model, does not specify running time, but the actual running time of the whole project is less than an hour. Second, Chang \& Li (\protect\hyperlink{ref-chang-2015}{2015}) believe researchers should ``set seeds and specify the random number generator.'' The information is very important and indeed included in our case. Third, they recommend authors to include the order of file execution, which is also detailed in the Readme file. Fourth, Chang \& Li (\protect\hyperlink{ref-chang-2015}{2015}) explicitly suggest that both raw data and transformed data should be incorporated. The data, as we stated earlier, was expected to be publicly accessible. Similar to the recommendations by Yale Law School, Chang \& Li (\protect\hyperlink{ref-chang-2015}{2015}) does not provide further suggestions to handle missing data files. Fifth, Chang \& Li (\protect\hyperlink{ref-chang-2015}{2015}) propose that the codes to produce results should be reported rather than ``taking parameters as given to produce results in tables and figures.'' This expectation is perfectly met in this case, because the author uses Sweave function (Rnw file) that allows codes to extract model results and produce a well-formatted LaTex document.

We can also examine the reproducibility of our case based on discipline-specific guidelines, which focus much on the reproducible methodology and analysis within an area. The most relevant guideline for our case study in education is the Companion Guidelines on Replication \& Reproducibility in Education Research (\protect\hyperlink{ref-national-science-foundation-2018}{National Science Foundation \& Institute of Education Sciences, U.S. Department of Education, 2018}), which highlights the aspects of reproducible proposal in education, transparency in intervention, open data access, open source codes and detailed information. In particular, this specific guideline stresses that the information should include a comparison with previous results, the methodology, the justification for data cleaning and funding information. All of the requirements are fulfilled and can be found in the original article (\protect\hyperlink{ref-skinner-2019B}{Skinner, 2019b}) and the GitHub repository (\protect\hyperlink{ref-skinner-2019B}{Skinner, 2019b}).

In summary, our case (\protect\hyperlink{ref-skinner-2019A}{Skinner, 2019a}, \protect\hyperlink{ref-skinner-2019B}{2019b}) in education satisfies most of the expectations perfectly in the guidelines/recommendations, except for the datasets (although the codes to download the datasets are attached), the information of software version and the expected running time of replication. Therefore we would expect the replication process would go relatively smoothly for our case.

\hypertarget{data}{%
\section{Data}\label{data}}

This paper uses the R statistical language (\protect\hyperlink{ref-r-2021}{R Core Team, 2021}) and \texttt{tidyverse} packages (\protect\hyperlink{ref-tidyverse-2019}{Wickham et al., 2019}). The database is connected to R using \texttt{DBI} package (\protect\hyperlink{ref-dbi-2021}{R Special Interest Group on Databases (R-SIG-DB) et al., 2021}). The report was made using RMarkdown (\protect\hyperlink{ref-rmd-2021}{Allaire et al., 2021}).

The data, in various formats, can be found in the Readme file under the data section in the GitHub repository (\protect\hyperlink{ref-skinner-2019B}{Skinner, 2019b}). A more detailed discussion can be found below.

\hypertarget{institution-data}{%
\subsection{Institution Data}\label{institution-data}}

The data are the same as the data in the original paper (\protect\hyperlink{ref-skinner-2019A}{Skinner, 2019a}) i.e.~data on the number of students who enroll in online courses were taken from the Integrated Postsecondary Education Data System (\protect\hyperlink{ref-skinner-2019A}{Skinner, 2019a}). Table 1 is the summary statistics table for the dataset i.e.~the institution sample (N is the number of institution in the specific year). All variables in Table 1 are incorporated in the models as covariates.

\begin{table}[!ht]
\centering
\caption{Descriptive statistics of the institution sample}
\label{tab:desc_ipeds}
\begin{tabularx}{\linewidth}{Xc}
\toprule
&Mean/(SD) \\
  \midrule
Total enrollment & 7725 \\ 
   & (7819) \\ 
  Some online enrollment & 1361 \\ 
   & (1536) \\ 
  Two year institution & 0.88 \\ 
   & (0.33) \\ 
  Has on-campus housing & 0.25 \\ 
   & (0.44) \\ 
  Non-white enrollment & 0.42 \\ 
   & (0.24) \\ 
  Women enrollment & 0.42 \\ 
   & (0.07) \\ 
  Pell grant recipients & 0.42 \\ 
   & (0.15) \\ 
  Part-time enrollment & 0.57 \\ 
   & (0.15) \\ 
  Aged 25 years and older & 0.37 \\ 
   & (0.11) \\ 
  2013 & 0.4 \\ 
   & (0.49) \\ 
  2014 & 0.3 \\ 
   & 0.46 \\ 
   \midrule
$N$ (2012) & 750 \\ 
  $N$ (2013) & 1003 \\ 
  $N$ (2014) & 741 \\ 
   \bottomrule

\multicolumn{2}{p{.98\linewidth}}
{\footnotesize{\itshape Notes.} 
Total enrollment and some online enrollment represent the average number of students rounded to nearest student. Other rows are proportions. Standard deviations are shown in parentheses. Schools included in the sample are public, open admissions postsecondary institutions that report at least one student who took some distance education courses. Although the table looks the same as the original paper [@skinner-2019A], the data were downloaded separately to reproduce this table.
}
\end{tabularx}
\end{table}

\hypertarget{demographics}{%
\subsection{Demographics}\label{demographics}}

The key demographic traits come from the United States Census Bureau (\protect\hyperlink{ref-skinner-2019A}{Skinner, 2019a}). A total of 3 datasets are used in the original paper (\protect\hyperlink{ref-skinner-2019A}{Skinner, 2019a}): the population, the employment by state and the employment by county. Though many variables are provided for the population dataset, only the year and the state population are useful in the report. Notably, there should be 50 states in the US but the datasets contain 51 states because the District of Columbia, a federal district, is treated as a state. Moreover, the top 10 most populated states have more than 50\% of the total population in the US. We illustrate the top 10 most populated states in Figure 1.

\begin{figure}[H]

{\centering \includegraphics[width=0.5\linewidth,height=0.5\textheight]{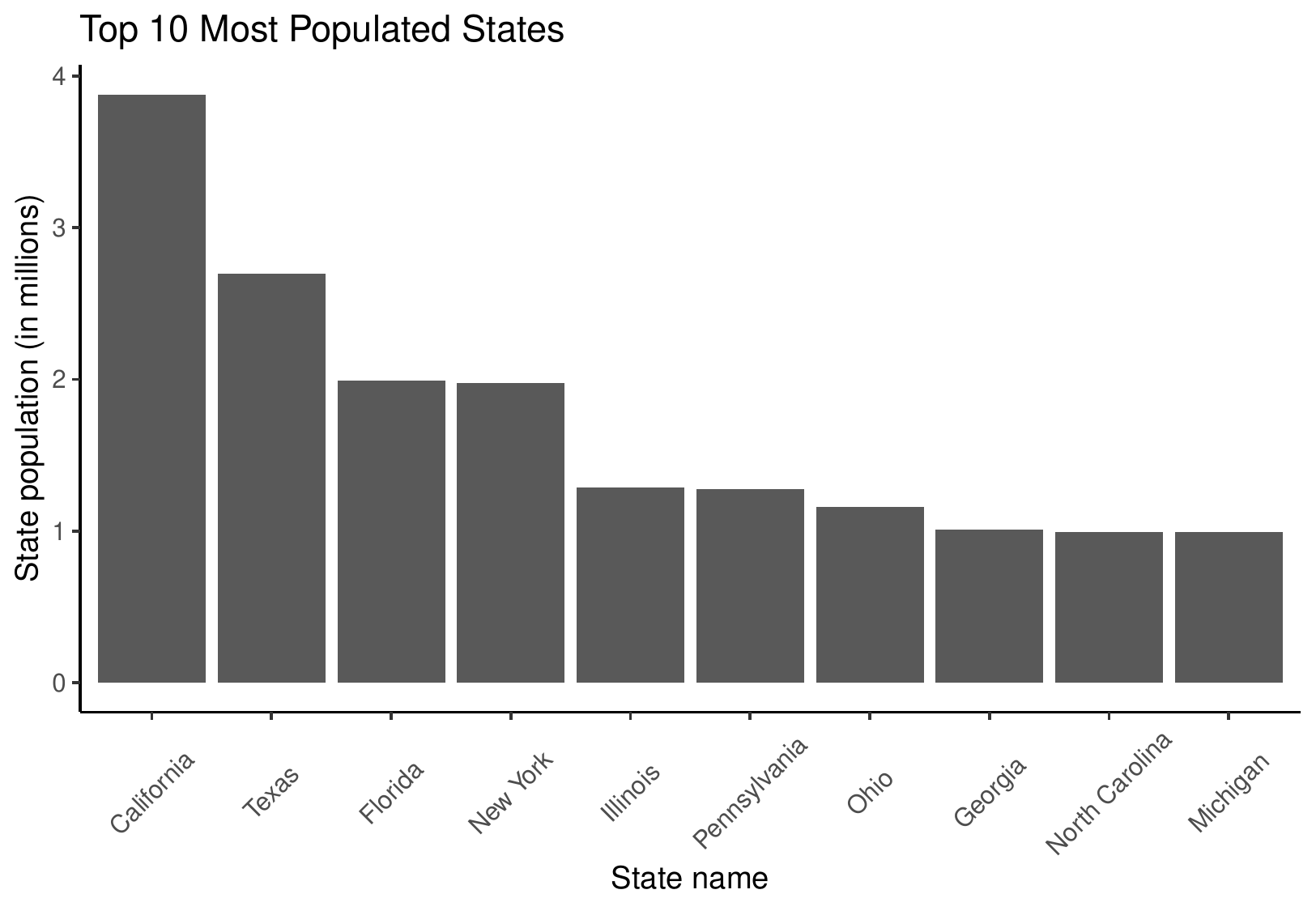} 

}

\caption{The 10 Largest States by Population in the US}\label{fig:demographics}
\end{figure}

\hypertarget{broadband-data}{%
\subsection{Broadband Data}\label{broadband-data}}

The key variables are the broadband data, that is, the download speed/upload speed/number of providers in each county/state in the US. All publicly available broadband data from 2012 to 2014 are from the State Broadband Initiative. A summary of the datasets can be found in Table 2.

\begin{table}[!ht]
\centering
\caption{Descriptive statistics of broadband measures}
\label{tab:desc_bb}
\begin{tabularx}{\linewidth}{Xccc}
\toprule
& 2012 & 2013 & 2014 \\
  \midrule
Download tier & 6.97 & 7.27 & 7.42 \\ 
   & (0.72) & (0.73) & (0.74) \\ 
  Upload tier & 4.43 & 4.66 & 4.94 \\ 
   & (0.81) & (0.89) & (0.94) \\ 
  Number of providers & 3.26 & 3.43 & 3.45 \\ 
   & (0.93) & (1.05) & (1) \\ 
   \bottomrule

\multicolumn{4}{p{.98\linewidth}}
{\footnotesize{\itshape Notes.} 
Values are the average of broadband measures assigned across all schools in the sample in a given year. Each school is given a value that is the population-distance-weighted average of surrounding measures (at the census block level). Download and upload speeds are reported in ordered categorical tiers from 2 to 11. Broadband data come from the National Broadband Map. Standard deviations are shown in parentheses. Although the table looks the same as the original paper [@skinner-2019A], the data were downloaded separately to reproduce this table.
}
\end{tabularx}
\end{table}

The visualization of Figure 2 shows that it is reasonable to assume the upload speed follows a normal distribution while the download speed is a right-skewed distribution. This indicates while some areas can enjoy the fast internet speed, the majority of areas still have slower download speed. Download speed is crucial for students when they are watching videos/downloading learning materials for online courses.

\begin{figure}

{\centering \includegraphics[width=400px,height=450px]{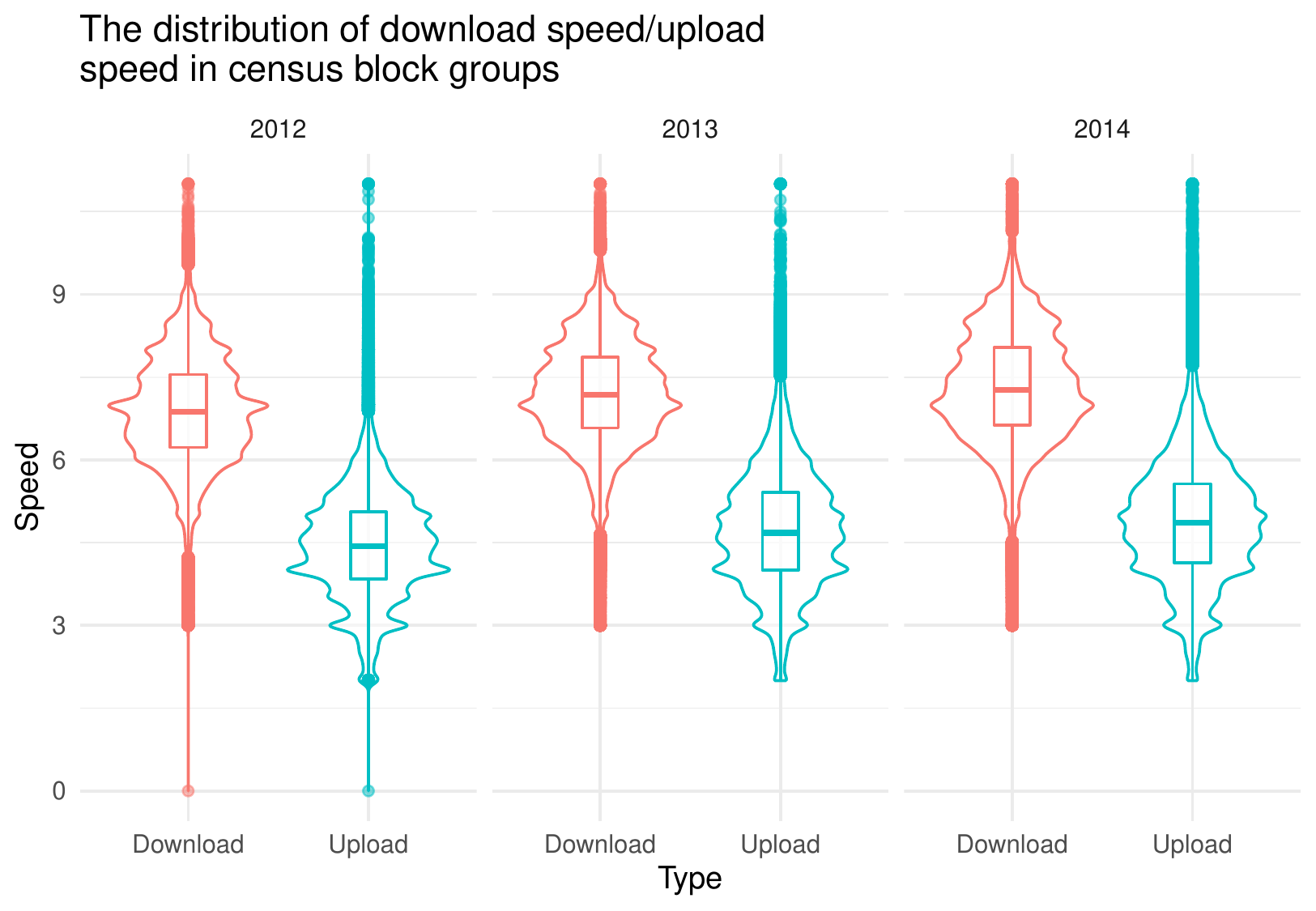} 

}

\caption{The distributions of upload speed and download speed in each census block group in the US from 2012 to 2014. The shape is approximately normal for the distribution of upload speed while the shape of the distribution of download speed is right-skewed.}\label{fig:dist-speed}
\end{figure}

The enrollment data are institution-based, while the broadband data are based on county/states. To account for the fact that students may have very different experiences given their distance from the institution and the fact that students may live in counties with various population sizes as shown in Figure 1, the author provides a re-weighting factor for the three types of broadband measure (upload speed, download speed, number of providers) of an institution. The detailed explanation can be found in the original paper (\protect\hyperlink{ref-skinner-2019A}{Skinner, 2019a}). The following equation shows the final product of a weighted broadband measure of an institution.

\begin{equation}
{wbroadband}_{s}=\sum_{c=1}^{C} \frac{\left(\begin{array}{c}
\frac{p o p_{c}}{\sum_{c=1}^{C} \text { pop }_{c}} \\
\end{array}\right)\left(\begin{array}{c}
\frac{i d_{s c}}{\sum_{c=1}^{C} i d_{s c}} \\
\end{array}\right) \cdot \text {broadband}_{c}}{\sum_{c=1}^{C}\left(\frac{p o p_{c}}{\sum_{c=1}^{C} p o p_{c}}\right)\left(\begin{array}{c}
\frac{i d_{s c}}{\sum_{c=1}^{C} i d_{s c}} 
\end{array}\right)}
\end{equation}

Weighted average broadband measures for each school, \(wbroadband_s\), is calculated based on the average broadband measure in each census block group \(wbroadband_c\). \(p o p_{c}\) is the population in one census block and \(id_{sc}\) is the inverse distance between the school and the census block i.e.~the larger the distance is, the smaller the inverse distance is.

\hypertarget{replicated-models}{%
\section{Replicated Models}\label{replicated-models}}

The replicated models are based on the models in the original paper (\protect\hyperlink{ref-skinner-2019A}{Skinner, 2019a}). Two types of model are used: single-level Bayesian linear regression model and multilevel Bayesian linear regression model.

The is a general formula of the single-level Bayesian linear regression models (\protect\hyperlink{ref-skinner-2019A}{Skinner, 2019a}):

\[
\log \left(y_{i}\right) \sim N\left(\alpha+\beta \text { Broadband }_{i}+X \gamma, \sigma_{y}^{2}\right)
\]

where \(y_{i}\) is the number of students who enroll in online courses; \(\alpha\) is a constant term representing the baseline number of students who enroll in online courses; \(\beta\) is the parameter of interest for Broadband \(_{i}\), the institution's assigned measure of broadband; and \(X\) is a matrix of covariate data values with \(\gamma\) as its corresponding vector of parameters. The composition of \(X\) can be found in the table under the Institution Data section. Logarithm is used to normalize the data. The variance is denoted as \(\sigma_{y}^{2}\). Logarithm also makes \(\beta\) represent the percentage change in the number of students who take online courses for each unit increase in the broadband measure of interest (\protect\hyperlink{ref-greene-2021}{Greene, 2021}).

The multilevel model is essentially a random-intercept model because each state \(j\) can have a different intercept \(\alpha_{j}\). Therefore, the multilevel model can be written as below:

\[
\begin{aligned}
\log \left(y_{i}\right) & \sim N\left(\alpha_{j}+\beta \text { Broadband }_{i}+X \gamma, \sigma_{y}^{2}\right) \\
\alpha_{j} & \sim N\left(\delta_{s} \operatorname{Region}_{s}+Z \psi, \sigma_{s}^{2}\right)
\end{aligned}
\]

\(\alpha_{j}\), is modeled using state-level covariates, \(Z\), and a region-specific intercept, \(\delta_{s} .\) In all models, all unknown parameters \(\left(\beta, \gamma, \psi, \delta_{s}, \sigma_{y}^{2}, \sigma_{s}^{2}\right)\) are given flat priors.

\hypertarget{results}{%
\section{Results}\label{results}}

\hypertarget{single-level-models}{%
\subsection{Single-level Models}\label{single-level-models}}

From left to right there are four models: (1) Model with download speed as broadband measure; (2) Model with upload speed as broadband measure; (3) Model with number of providers as broadband measure; (4) Full model with all three types of broadband measures.

The parameters from the Table 3 align with most of the reported parameters in the original paper except that a few numbers differ in the thousandths digit. This means we may be able to arrive at a similar conclusion. First, the estimate of the effect of year variable, \(\beta_{2013}=0.103\) and \(\beta_{2014}=0.15\), show positive growth in the number of students in online courses with growth rate around 11\% to 16\% since 2012. These findings are exactly the same as the original paper (\protect\hyperlink{ref-skinner-2019A}{Skinner, 2019a}).

Second, there is a positive relationship between online course enrollment in an institution and the download speed (\(\beta_{download}= 0.335\) with 95\%CI{[}-0.019,0.681{]} in the first model and \(\beta_{download}= 0.405\) 95\%CI{[}-0.011,0.811{]} in the full model) as well as the number of providers (\(\beta_{provider}= 0.008\), 95\%CI{[}-0.104,0.119{]} in the third model and \(\beta_{provider}= 0.029\) with 95\%CI{[}-0.084,0.141{]} in the full model). Moreover, two-year institutions on average have around 6\% more students in online courses compared to four-year institutions assuming other variables are constant. These variables, however, may have no effects on online course enrollment and the observed effects are only due to random chances because the 95\% credible intervals of the estimates of these parameters contain 0.

Furthermore, proportions of women, proportions of students of color, and proportions of part-time students are associated negatively with proportions of students taking online courses. This result is explained with ``non-traditional student'' designation (\protect\hyperlink{ref-snyder-2016}{Snyder et al., 2016}), I believe these results also imply that the popular courses with large class size are still dominated by a specific group of students.

The intercept \(\alpha\) is the expected log number of students enrolled in the online courses (\protect\hyperlink{ref-skinner-2019A}{Skinner, 2019a}), but the intercept is only useful when all other independent variables are 0. Based on the models, since broadband measures and the other covariates cannot be 0, \(\alpha\) has no practical meaning.

\hypertarget{multilevel-models}{%
\subsection{Multilevel Models}\label{multilevel-models}}

The results on Table 4 agree with the original paper except that some numbers differ in the thousandths digit. The table also implies that the conclusion from multilevel models is similar to that from the single-level models. Note that the Bayesian point estimate for download speed in the full model is positive (\(\beta_{download} =0.537\)) and with a 95\% credible interval that does not cover zero {[}0.12, 0.947{]}.

\hypertarget{pain-points}{%
\section{Pain Points}\label{pain-points}}

The actual replication took me a week to replicate the same results on Linux. In the following sections, I am going to discuss four main challenges of replication of this specific case: 1. The implementation of \texttt{make} utility; 2. The access to public data; 3. The problem of version control; 4. Some difficulties with R.

\hypertarget{gnu-make-utility-and-cut-utility}{%
\subsection{GNU make Utility and Cut Utility}\label{gnu-make-utility-and-cut-utility}}

GNU make is a standard implementation of Make for Linux and macOS. Therefore, in order to run the one-step replication which is written on MacOS, one should be equipped with Unix or Unix-like systems. In order to test the functionality, Ubuntu 20.04 was installed along with other required software (R, SQLite, LaTex) on Windows 10. Note that Linux is never expected to be taught to many statistics students, so it took me half a day to learn the basic syntax.

Even if one has prior knowledge in different operating systems, the shell scripts may still produce errors due to subtle differences in the usage of functions. These differences may be hard to discern for statisticians like me, and will cause actual problems when replicating with original source codes. For instance, the \texttt{cut} function in \texttt{Linux} no longer allows \texttt{cut\ -f4\ -f5} but \texttt{cut\ -f4,5} in the latest version. Therefore, when running the bash file to extract a substring from a data file name, the makefile will throw an error in the Linux environment. These errors may be tricky to interpret depending on the researchers' background. In general, this type of situation can only be solved when one has experience with shell scripts in different operating systems for some years or one has the ability to debug efficiently.

\newpage

\begin{table}[!ht]
\centering
\caption{Single level Bayesian regressions of log number of students who enrolled in some distance education courses on broadband measures.}
\label{tab:sl_normal_full}
\begin{tabularx}{\linewidth}{Xcccc}
\toprule
& (1) & (2) & (3) & (4) \\
  \midrule
Download speed & 0.335 &  &  & 0.405 \\ 
   & [-0.019,0.681] &  &  & [-0.011,0.811] \\ 
  Download speed$^2$ & -0.025 &  &  & -0.027 \\ 
   & [-0.049,0] &  &  & [-0.055,0.003] \\ 
  Upload speed &  & -0.076 &  & -0.21 \\ 
   &  & [-0.27,0.117] &  & [-0.442,0.019] \\ 
  Upload speed$^2$ &  & 0.003 &  & 0.015 \\ 
   &  & [-0.017,0.023] &  & [-0.007,0.038] \\ 
  \# Providers &  &  & 0.008 & 0.029 \\ 
   &  &  & [-0.104,0.119] & [-0.084,0.141] \\ 
  \# Providers$^2$ &  &  & -0.003 & -0.004 \\ 
   &  &  & [-0.016,0.011] & [-0.017,0.01] \\ 
  Two year institution & 0.058 & 0.067 & 0.065 & 0.056 \\ 
   & [-0.017,0.131] & [-0.007,0.14] & [-0.015,0.142] & [-0.018,0.132] \\ 
  Has on-campus housing & -0.049 & -0.066 & -0.048 & -0.063 \\ 
   & [-0.118,0.02] & [-0.138,0.004] & [-0.12,0.02] & [-0.138,0.011] \\ 
  $log$(Total enrollment) & 1.154 & 1.156 & 1.154 & 1.157 \\ 
   & [1.116,1.191] & [1.119,1.194] & [1.118,1.19] & [1.12,1.193] \\ 
  Prop. non-white & -0.633 & -0.629 & -0.635 & -0.612 \\ 
   & [-0.759,-0.511] & [-0.742,-0.513] & [-0.753,-0.521] & [-0.731,-0.494] \\ 
  Prop. women & -2.218 & -2.196 & -2.231 & -2.199 \\ 
   & [-2.636,-1.795] & [-2.613,-1.788] & [-2.673,-1.808] & [-2.619,-1.778] \\ 
  Prop. Pell grant & 0.614 & 0.569 & 0.608 & 0.577 \\ 
   & [0.404,0.829] & [0.357,0.782] & [0.396,0.826] & [0.363,0.797] \\ 
  Prop. part-time & -0.476 & -0.492 & -0.464 & -0.489 \\ 
   & [-0.703,-0.25] & [-0.72,-0.261] & [-0.702,-0.236] & [-0.722,-0.257] \\ 
  Prop. 25 years and older & 0.394 & 0.402 & 0.406 & 0.417 \\ 
   & [0.141,0.651] & [0.16,0.646] & [0.161,0.654] & [0.172,0.664] \\ 
  $log$(Pop. density) & -0.091 & -0.084 & -0.087 & -0.087 \\ 
   & [-0.117,-0.065] & [-0.11,-0.058] & [-0.113,-0.06] & [-0.112,-0.061] \\ 
  2013 & 0.103 & 0.109 & 0.1 & 0.105 \\ 
   & [0.046,0.158] & [0.052,0.168] & [0.042,0.157] & [0.046,0.161] \\ 
  2014 & 0.15 & 0.162 & 0.142 & 0.158 \\ 
   & [0.087,0.213] & [0.099,0.228] & [0.08,0.203] & [0.097,0.222] \\ 
  (Intercept) & 6.726 & 6.726 & 6.726 & 6.726 \\ 
   & [6.703,6.749] & [6.703,6.749] & [6.703,6.75] & [6.702,6.75] \\ 
   \midrule
Unique institutions & 1017 & 1017 & 1017 & 1017 \\ 
  $N$ & 2494 & 2494 & 2494 & 2494 \\ 
   \bottomrule

\multicolumn{5}{p{.98\linewidth}}
{\footnotesize{\itshape Notes.} 
Bayesian point estimates represent posterior mean values. Values in the square brackets are 95\% credible intervals. Covariates not reported include indicators for USDA urban/rural community codes. Parameter distributions in each model are the combination of four independent MCMC chains of 1000 draws each (with 1000 initial draws discarded as burn-in) for a total of 4000 draws. All models were estimated using the Stan NUTS sampler with a normal likelihood sampling statement. The outcome measure in all models is the log number of students at each institution who enrolled in some distance education courses. Although the table looks similar to the original paper [@skinner-2019A], the data were downloaded separately and models were compiled on my local computer to reproduce this table. Except for a small difference in the decimal places, the results looks the same as the original one.
}
\end{tabularx}
\end{table}

\newpage

\begin{table}[!ht]
\centering
\caption{Varying intercept Bayesian regressions of log number of students who enrolled in some distance education courses on broadband measures}
\label{tab:vi_normal_full}
\begin{tabularx}{\linewidth}{Xcccc}
\toprule
& (1) & (2) & (3) & (4) \\
  \midrule
Download speed & 0.329 &  &  & 0.537 \\ 
   & [-0.033,0.672] &  &  & [0.12,0.947] \\ 
  Download speed$^2$ & -0.025 &  &  & -0.039 \\ 
   & [-0.048,0.001] &  &  & [-0.068,-0.01] \\ 
  Upload speed &  & -0.105 &  & -0.206 \\ 
   &  & [-0.305,0.096] &  & [-0.444,0.023] \\ 
  Upload speed$^2$ &  & 0.008 &  & 0.02 \\ 
   &  & [-0.012,0.028] &  & [-0.002,0.044] \\ 
  \# Providers &  &  & -0.044 & -0.059 \\ 
   &  &  & [-0.153,0.063] & [-0.175,0.053] \\ 
  \# Providers$^2$ &  &  & 0.003 & 0.004 \\ 
   &  &  & [-0.009,0.016] & [-0.009,0.018] \\ 
  Two year institution & 0.087 & 0.099 & 0.097 & 0.091 \\ 
   & [-0.002,0.174] & [0.01,0.186] & [0.009,0.186] & [0.005,0.179] \\ 
  Has on-campus housing & -0.017 & -0.025 & -0.024 & -0.024 \\ 
   & [-0.088,0.055] & [-0.099,0.049] & [-0.096,0.047] & [-0.101,0.053] \\ 
  $log$(Total enrollment) & 1.132 & 1.131 & 1.133 & 1.135 \\ 
   & [1.096,1.171] & [1.095,1.169] & [1.096,1.171] & [1.097,1.173] \\ 
  Prop. non-white & -0.781 & -0.776 & -0.773 & -0.787 \\ 
   & [-0.935,-0.623] & [-0.934,-0.623] & [-0.933,-0.612] & [-0.948,-0.628] \\ 
  Prop. women & -2.11 & -2.113 & -2.113 & -2.134 \\ 
   & [-2.517,-1.689] & [-2.514,-1.696] & [-2.528,-1.727] & [-2.524,-1.71] \\ 
  Prop. Pell grant & 0.462 & 0.443 & 0.433 & 0.447 \\ 
   & [0.208,0.716] & [0.191,0.685] & [0.19,0.688] & [0.191,0.691] \\ 
  Prop. part-time & -0.849 & -0.84 & -0.853 & -0.856 \\ 
   & [-1.111,-0.597] & [-1.094,-0.576] & [-1.111,-0.609] & [-1.115,-0.597] \\ 
  Prop. 25 years and older & 0.263 & 0.258 & 0.269 & 0.275 \\ 
   & [-0.004,0.529] & [-0.009,0.522] & [0.008,0.535] & [0.008,0.546] \\ 
  $log$(Pop. density) & -0.065 & -0.058 & -0.059 & -0.059 \\ 
   & [-0.096,-0.037] & [-0.088,-0.027] & [-0.09,-0.03] & [-0.09,-0.028] \\ 
  2013 & 0.099 & 0.1 & 0.095 & 0.105 \\ 
   & [0.045,0.152] & [0.046,0.154] & [0.043,0.148] & [0.051,0.16] \\ 
  2014 & 0.158 & 0.16 & 0.15 & 0.167 \\ 
   & [0.099,0.218] & [0.097,0.22] & [0.093,0.209] & [0.104,0.227] \\ 
   \midrule
Unique institutions & 1017 & 1017 & 1017 & 1017 \\ 
  $N$ & 2494 & 2494 & 2494 & 2494 \\ 
   \bottomrule

\multicolumn{5}{p{.98\linewidth}}
{\footnotesize{\itshape Notes.} 
Bayesian point estimates represent posterior mean values. Values in the square brackets are 95\% credible intervals. Intercepts (not reported) were allowed to vary at the state level. First level covariates not reported are indicators for USDA urban/rural community codes. Second level covariates include state unemployment rate, statewide average appropriations per FTE student, the proportion of public open admissions institutions in the state that are two-year institutions, and a population-weighted measure of the average distance to the nearest open admissions institution in the state. Parameter distributions in each model are the combination of four independent MCMC chains of 1000 draws each (with 1000 initial draws discarded as burn-in) for a total of 4000 draws. All models were estimated using the Stan NUTS sampler with a normal likelihood sampling statement. The outcome measure in all models is the log number of students at each institution who enrolled in some distance education courses. Although the table looks similar to the original paper [@skinner-2019A], the data were downloaded separately and models were compiled on my local computer to reproduce this table. Except for a small difference in the decimal places, the results looks the same as the original one.
}
\end{tabularx}
\end{table}

\hypertarget{data-accessibility}{%
\subsection{Data Accessibility}\label{data-accessibility}}

Public data can sometimes be a timed bomb for reproducible research. The accessibility of data is a real concern of scientific replication for three reasons: the data files may be moved, the access to data files may be restricted and the data file may be altered without notifying researchers. In our replication, the first two problems appeared to be obstacles.

The first challenge is related to data storage i.e.~the data files may be removed or the website may change its link to the datasets. During my replication process, State\_by\_State\_Wave\_Charts\_FY15\_0.xlsx was missing from the original link and it required some effort to find the new url link to the data. However, the new url had no file extension, so the scripts for downloading data did not work for this file. The most efficient solution, in this case, was to download the data manually to the expected directory and change the name accordingly. This worked just fine for a single missing dataset. What if there are many public datasets that are not available due to change in domain name or permanent removal of data files?

Second, data may not be accessible from certain regions. For instance, I found that the US broadband data was not accessible from the University of Toronto (UofT) internet, so the makefile would throw an error when downloading data. The solution was to connect to another WIFI, but this would inevitably increase the cost of a replication project. Furthermore, the data may limit the number of download attempts. In my case, the computer once stopped downloading some files for unknown reasons (possibly because of the restricted number of download attempts allowed for each IP). In the above three situations, the obtained data can be either incomplete or missing.

Third, there is no guarantee that the datasets are still the same after the publication of a research paper. Ideally, any modifications on the public datasets will be recorded on their websites. However, if there is no version number for the datasets, researchers cannot be sure that the public data files downloaded now are exactly the same as the ones used in the original research projects. This problem can only be checked if the authors of the paper recorded the hash codes, a file identifier, of the data files in Readme files.

In general, none of the existing guidelines in the last section provide satisfactory solutions to the problem of accessible public data and consistent datasets files. In the Next Steps section, we advocate two recommendations regarding the potential issues of public data accessibility.

\hypertarget{version-control}{%
\subsection{Version Control}\label{version-control}}

The obscurity in the version number of software and R packages also adds some difficulties in reproducing the final report from the Sweave document. This is because the latest packages will always be installed by default. This, however, can actually lead to problems. In this case study, figures about model results are based on the information extracted from the stanfit objects in R. In particular, \texttt{tidy} from the \texttt{broom} package is used to convert the fitted objects into a tidy dataframe. However, these methods are already moved to the \texttt{broom.mixed} package in 2021, so the output pdf will demonstrate this error: No tidy method for objects of class stanfit. The solution in this case is rather simple: change the package to broomExtra, which contains functions from both broom and broom.mixed. Ideally, this situation can be avoided if we know the versions of the packages and download the corresponding versions of the R packages for replication. However, it is possible that the old R packages are not compatible with the latest R or the operating system. This requires extra effort in choosing the appropriate versions of applications.

\hypertarget{some-difficulties-with-r}{%
\subsection{Some Difficulties with R}\label{some-difficulties-with-r}}

RStudio works in various ways on different operating systems. When downloading R packages, it is important to include \texttt{dependencies\ =\ True}, otherwise users may fail to install \texttt{devtools}, \texttt{V8} and \texttt{rstan} on Linux because of missing dependencies. In such cases, one has to open R to manually install these packages. I failed a couple of times when dealing with installing \texttt{R} packages. Windows RStudio allows automatic installation of the dependencies, but this works differently for R in Linux.

Furthermore, the generation of our final report requires R to compile the Sweave document into pdf. However, R returned different types of errors related to LaTex in a Linux environment. Even if we switch back to Windows and try to compile the file via Windows RStudio, similar errors occurred. After a long day of debugging, from reinstalling R and MikTex to switching between operating systems, we found a final two-step solution with Knitr package. In particular, Knitr converted the Sweave document to a Tex file first and then knitted the Tex file to pdf. This method finally worked, but it also raised many problems regarding the reproducibility of R with Sweave and MikTex.

\hypertarget{discussion}{%
\section{Discussion}\label{discussion}}

\hypertarget{a-reflection-on-undergraduate-statistics-education}{%
\subsection{A Reflection on Undergraduate Statistics Education}\label{a-reflection-on-undergraduate-statistics-education}}

This experience has offered us an opportunity to glimpse what programming skills and statistical knowledge are required for a published research project. More importantly, it reminds us what undergraduate education should offer to students in order for them to complete such a simple replication. In this section, based on our replication experience, we am going to discuss an important yet usually missing aspect in undergraduate statistics education: the database management.

\hypertarget{the-teaching-of-database-management}{%
\subsubsection{The Teaching of Database Management}\label{the-teaching-of-database-management}}

In our project, we used basic SQL commands to clean and join new tables. I have learnt it before by myself, so it is not difficult for me. However, most undergraduate statistics programs do not require the teaching of SQL programming/database management. According to a recent survey, R and Python still dominate the undergraduate curriculum and SQL is rarely taught (\protect\hyperlink{ref-schwab-mccoy-2021}{Schwab-McCoy et al., 2021}). Nonetheless, the teaching of database management (SQL/NoSQL) has been listed as one of the expectations in a wide range of curriculum guidelines for undergraduate statistics education (\protect\hyperlink{ref-asa-2014}{American Statistical Association Undergraduate Guidelines Workgroup, 2014}; \protect\hyperlink{ref-de-veaux-2017}{De Veaux et al., 2017}; \protect\hyperlink{ref-nolan-2010}{Nolan \& Temple Lang, 2010}). Moreover, SQL and NoSQL rank 6 and 7 in the ranking of Most In-Demand Programming Languages for 2021, according to a study by University of California, Berkeley (\protect\hyperlink{ref-ucb-2021}{2021}). Though the ranking is based on Stack Overflow's developer surveys in 2019 and 2020, it still shows the importance of query languages in statistics-related fields.

Nonetheless, the approaches to teach databases can vary. SQL is often combined with R tidyverse instead of SQL programming language itself in actual projects. For example, Broatch et al. (\protect\hyperlink{ref-broatch-2019}{2019}) shows the workflow to import db into R and clean data with dplyr. In another collaborative project (\protect\hyperlink{ref-loy-2019}{Loy et al., 2019}), the core concepts of relational database (SQL) were introduced, but the data management was done with tidyverse again. There are two major limitations of such approaches. First, students do not have the opportunity to practice SQL programming, which would possibly be a requirement if students would like to enter the workplace after graduation. Moreover, in these projects, databases are treated simply as larger csv/json files in the workflow. Brunner \& Kim (\protect\hyperlink{ref-brunner-2016}{2016}) suggests an approach that takes both the advantages of prior knowledge in Python as well as the benefits of SQL programming in command lines. My self-learning experience illustrates that R and SQL programming can also be taught simultaneously. During summer 2021, I taught myself SQL by comparing the basic commands in SQL with R commands (\protect\hyperlink{ref-zheng-2021}{Zheng, 2021}). The series of notes can be found on my personal website and the actual learning lasted around 1 month. We believe teaching SQL in undergraduate can even smoothen the learning curve for students who are going to use SQL in their future careers.

\hypertarget{next-steps-for-reproducibility-guidelines}{%
\subsection{Next Steps for Reproducibility Guidelines}\label{next-steps-for-reproducibility-guidelines}}

The pain points section elaborates some hidden challenges to replicate a case study that satisfies almost all expectations of guidelines. Instead of focusing on the discussion of methodology and analysis, we want to emphasize the real technical issues and the lessons from a statistician's perspective. Despite the issues inherent in R, we want to draw the attention to the importance of new guidelines that take care of the data access problems and the version control. Clearly, existing guidelines of reproducible research do not capture these two important aspects of reproducible research.

If we anticipate such difficulties in long-term reproducibility due to the accessibility of public data, what can we do? We recommend at least three ways for the issue. First, the author can record the Hash codes so that if researchers acquire the datasets elsewhere, they can check if the data is the same as the original paper. A checker can be written via shell script. Second, the author can save a small subset of the data if permitted and run the program on this subset. Then the author can present the codes, the sample data itself and the figures and graph related to model results in the appendix. It ensures that scholars can keep track of the workflow when they try to reproduce the results step-by-step. If this is not allowed, the author can generate some random data and report the data generation scripts and the generated data as in the second recommendation. Then future researchers are able to at least evaluate the reproducibility of the workflow based on the source codes. We believe the last recommendation will serve well as a back-up plan in case some datasets are permanently removed.

\newpage

\hypertarget{references}{%
\section*{References}\label{references}}
\addcontentsline{toc}{section}{References}

\hypertarget{refs}{}
\begin{CSLReferences}{1}{0}
\leavevmode\vadjust pre{\hypertarget{ref-rmd-2021}{}}%
Allaire, J., Xie, Y., McPherson, J., Luraschi, J., Ushey, K., Atkins, A., Wickham, H., Cheng, J., Chang, W., \& Iannone, R. (2021). \emph{Rmarkdown: Dynamic documents for r}. \url{https://github.com/rstudio/rmarkdown}

\leavevmode\vadjust pre{\hypertarget{ref-asa-2014}{}}%
American Statistical Association Undergraduate Guidelines Workgroup. (2014). \emph{{Curriculum Guidelines for Undergraduate Programs in Statistical Science}}. \url{https://www.amstat.org/asa/files/pdfs/EDU-guidelines2014-11-15.pdf}

\leavevmode\vadjust pre{\hypertarget{ref-broatch-2019}{}}%
Broatch, J. E., Dietrich, S., \& Goelman, D. (2019). {Introducing Data Science Techniques by Connecting Database Concepts and dplyr}. \emph{Journal of Statistics Education}, \emph{27}(3), 147--153. \url{https://doi.org/10.1080/10691898.2019.1647768}

\leavevmode\vadjust pre{\hypertarget{ref-brunner-2016}{}}%
Brunner, R. J., \& Kim, E. J. (2016). {Teaching Data Science}. \emph{Procedia Computer Science}, \emph{80}, 1947--1956. \url{https://doi.org/10.1016/j.procs.2016.05.513}

\leavevmode\vadjust pre{\hypertarget{ref-chang-2015}{}}%
Chang, A. C., \& Li, P. (2015). {Is Economics Research Replicable? Sixty Published Papers from Thirteen Journals Say "Usually Not"}. \emph{Finance and Economics Discussion Series}, \emph{2015}(83), 1--26. \url{https://doi.org/10.17016/feds.2015.083}

\leavevmode\vadjust pre{\hypertarget{ref-de-veaux-2017}{}}%
De Veaux, R. D., Agarwal, M., Averett, M., Baumer, B. S., Bray, A., Bressoud, T. C., Bryant, L., Cheng, L. Z., Francis, A., Gould, R., Kim, A. Y., Kretchmar, M., Lu, Q., Moskol, A., Nolan, D., Pelayo, R., Raleigh, S., Sethi, R. J., Sondjaja, M., \ldots{} Ye, P. (2017). {Curriculum Guidelines for Undergraduate Programs in Data Science}. \emph{Annual Review of Statistics and Its Application}, \emph{4}(1), 15--30. \url{https://doi.org/10.1146/annurev-statistics-060116-053930}

\leavevmode\vadjust pre{\hypertarget{ref-greene-2021}{}}%
Greene, W. (2021). \emph{{Econometric Analysis Global Edition}} (8th edition). Pearson-prentice Hall.

\leavevmode\vadjust pre{\hypertarget{ref-loy-2019}{}}%
Loy, A., Kuiper, S., \& Chihara, L. (2019). {Supporting Data Science in the Statistics Curriculum}. \emph{Journal of Statistics Education}, \emph{27}(1), 2--11. \url{https://doi.org/10.1080/10691898.2018.1564638}

\leavevmode\vadjust pre{\hypertarget{ref-national-science-foundation-2018}{}}%
National Science Foundation, \& Institute of Education Sciences, U.S. Department of Education. (2018). \emph{{Companion Guidelines on Replication \& Reproducibility in Education Research}}. Institute of Education Sciences. \url{https://ies.ed.gov/pdf/CompanionGuidelinesReplicationReproducibility.pdf}

\leavevmode\vadjust pre{\hypertarget{ref-nolan-2010}{}}%
Nolan, D., \& Temple Lang, D. (2010). {Computing in the Statistics Curricula}. \emph{The American Statistician}, \emph{64}(2), 97--107. \url{https://doi.org/10.1198/tast.2010.09132}

\leavevmode\vadjust pre{\hypertarget{ref-r-2021}{}}%
R Core Team. (2021). \emph{R: A language and environment for statistical computing}. R Foundation for Statistical Computing. \url{https://www.R-project.org/}

\leavevmode\vadjust pre{\hypertarget{ref-dbi-2021}{}}%
R Special Interest Group on Databases (R-SIG-DB), Wickham, H., \& Müller, K. (2021). \emph{DBI: R database interface}. \url{https://CRAN.R-project.org/package=DBI}

\leavevmode\vadjust pre{\hypertarget{ref-schwab-mccoy-2021}{}}%
Schwab-McCoy, A., Baker, C. M., \& Gasper, R. E. (2021). {Data Science in 2020: Computing, Curricula, and Challenges for the Next 10 Years}. \emph{Journal of Statistics and Data Science Education}, \emph{29}(sup1), S40--S50. \url{https://doi.org/10.1080/10691898.2020.1851159}

\leavevmode\vadjust pre{\hypertarget{ref-skinner-2019A}{}}%
Skinner, B. T. (2019a). \emph{{GitHub: Replication files for Skinner (2019) Making the connection: Broadband access and online course enrollment at public open admissions institutions.}} \url{https://github.com/btskinner/oa_online_broadband_rep}

\leavevmode\vadjust pre{\hypertarget{ref-skinner-2019B}{}}%
Skinner, B. T. (2019b). {Making the Connection: Broadband Access and Online Course Enrollment at Public Open Admissions Institutions}. \emph{Research in Higher Education}, \emph{60}(7), 960--999. \url{https://doi.org/10.1007/s11162-018-9539-6}

\leavevmode\vadjust pre{\hypertarget{ref-snyder-2016}{}}%
Snyder, T. D., Brey, C. de, \& Dillow, S. A. (2016). \emph{{Digest of Education Statistics 2015 (NCES 2016-014)}}. \url{https://nces.ed.gov/pubs2016/2016014.pdf}

\leavevmode\vadjust pre{\hypertarget{ref-tetens-2016}{}}%
Tetens, H. (2016). {Reproducibility, Objectivity, Invariance}. \emph{Reproducibility: Principles, Problems, Practices, and Prospects}, 13--20. \url{https://doi.org/10.1002/9781118865064.ch1}

\leavevmode\vadjust pre{\hypertarget{ref-ucb-2021}{}}%
University of California, Berkeley. (2021). \emph{{11 Most In-Demand Programming Languages in 2021}}. \url{https://bootcamp.berkeley.edu/blog/most-in-demand-programming-languages/}

\leavevmode\vadjust pre{\hypertarget{ref-wicherts-2006}{}}%
Wicherts, J. M., Borsboom, D., Kats, J., \& Molenaar, D. (2006). {The poor availability of psychological research data for reanalysis.} \emph{American Psychologist}, \emph{61}(7), 726--728. \url{https://doi.org/10.1037/0003-066x.61.7.726}

\leavevmode\vadjust pre{\hypertarget{ref-tidyverse-2019}{}}%
Wickham, H., Averick, M., Bryan, J., Chang, W., McGowan, L. D., François, R., Grolemund, G., Hayes, A., Henry, L., Hester, J., Kuhn, M., Pedersen, T. L., Miller, E., Bache, S. M., Müller, K., Ooms, J., Robinson, D., Seidel, D. P., Spinu, V., \ldots{} Yutani, H. (2019). Welcome to the {tidyverse}. \emph{Journal of Open Source Software}, \emph{4}(43), 1686. \url{https://doi.org/10.21105/joss.01686}

\leavevmode\vadjust pre{\hypertarget{ref-readxl-2019}{}}%
Wickham, H., \& Bryan, J. (2019). \emph{Readxl: Read excel files}. \url{https://CRAN.R-project.org/package=readxl}

\leavevmode\vadjust pre{\hypertarget{ref-yale-2010}{}}%
Yale, S. (2010). {Reproducible Research}. \emph{Computing in Science \& Engineering}, \emph{12}(5), 8--13. \url{https://doi.org/10.1109/mcse.2010.113}

\leavevmode\vadjust pre{\hypertarget{ref-zheng-2021}{}}%
Zheng, S. (2021). \emph{{Learning SQL Notes}}. \url{http://siqi-zheng.rbind.io/post/2021-05-26-sql-notes-1/}

\end{CSLReferences}

\end{document}